\documentclass[english
]{acm_sen_article}

\usepackage{cleveref}
\usepackage{tabularx}
\usepackage{ctable}
\usepackage{url}
\usepackage{paralist}

\newcommand{\furl}[1]{\footnote{See \url{#1}.}}
\newcommand{\fontfix}{\LARGE\sffamily\bfseries}

\begin{document}
\title{
  SCAV'18: Report of the 2{\fontfix $^{nd}$} International Workshop on\\
  Safe Control of Autonomous Vehicles%
  {\fontfix\thanks{Preprint submitted to ACM SIGSOFT SEN.
  }}
}

\author{
  Mario Gleirscher\thanks{Supported by the
    Deutsche Forschungsgemeinschaft (DFG) under the Grant no. GL 915/1-1.}\\
  \affaddr{University of York, UK}\\
  \email{mario.gleirscher@york.ac.uk} \and
  Stefan Kugele\\
  \affaddr{Technical University of Munich, Germany}\\
  \email{stefan.kugele@tum.de} \and
  Sven Linker\\
  \affaddr{University of Liverpool, UK}\\
  \email{s.linker@liverpool.ac.uk} }

\maketitle

\sloppy

\begin{abstract}
  This report summarizes the discussions, open
  issues, take-away messages, and conclusions of the 2$^{nd}$ SCAV workshop.
\end{abstract}


\section{Organizational Summary}
\label{sec:workshop-summary}

\paragraph{Retrospect of SCAV'17}

Our main goal was to frame the not so clearly defined problem of
\emph{safe autonomous vehicle (AV) control}.  From the 1$^{st}$ SCAV
workshop~\cite{Gleirscher2017-SCAV-proceedings}, we were able to
determine key challenges as summarized in a report in 
\cite{Gleirscher2017-SCAV-report}.

\paragraph{SCAV'18}

This year, we focused on \textbf{applied formal verification of AV
  controllers} as well as on discussing their fault-tolerance and
assurance.  We give an overview of the achievements and identify
further issues in the verification of AV control.

Our blind peer review process included 3 to 4 reviews per paper,
including occasional shepherding.  We were able to accept 6 out of 8
submissions.  The program is outlined in \Cref{tab:program} and
documented in \cite{Gleirscher2018-SCAV-proceedings}.
For SCAV'18,\furl{https://scav.in.tum.de} we again asked paper authors
to act as discussants for initiating the question \& answer sessions
after each talk.  This way, the discussions appeared more lively
because authors get to know each others' works more closely.

\emph{Khalil Ghorbal} from the INRIA
group\furl{https://team.inria.fr/hycomes} for \emph{Hybrid Modeling
  and Contract-Based Design for Multi-physics Embedded Systems}
gave an excellent keynote on the state-of-the-art of invariant
checking and generation for non-linear systems, pointing at
current challenges.  His presentation on \emph{Simulating and
  Verifying Cyber-Physical Systems (CPSs): Current Challenges and Novel
  Research Directions} shed light on steps to take towards
industrial-scale CPSs and provided a smooth transition to our morning
session.

\paragraph{Acknowledgments}
\label{sec:thanks-all-contr}

Latest at this point, we would like to thank all the authors and
presenters\footnote{See the list of authors in the workshop proceedings
  \cite{Gleirscher2018-SCAV-proceedings}.} for putting great effort in
their talks as well as our discussants and the audience for posing the
right questions and pointing out important issues.  Furthermore, our
program committee\footnote{See the list of members and sub-reviewers in
  \cite{Gleirscher2018-SCAV-proceedings}.} deserves sincere gratitude
for their great standards and careful work.  Finally, we thank the
CPSWeek 2018 organizers for a high-quality scientific event in a
beautiful place full of history: Porto in Portugal.

\begin{table}[t]
  \caption{Outlined workshop program
    \label{tab:program}}
  \begin{tabularx}{\columnwidth}{lX}
    \toprule
    9:15 --  10:00 & \textbf{Keynote} by Khalil Ghorbal
    \\\midrule
    10:30 -- 12:30 & \textbf{Morning session:\newline Modeling,
      verification, simulation} \newline (3 talks)
    \\\midrule
    14:00 -- 15:30 & \textbf{Afternoon session:\newline Failure
      analysis, tolerance, safety argumentation} \newline (3 talks)
    \\\midrule
    16:40 -- 17:30 & \textbf{Discussion and tool demo}\\
    \bottomrule
  \end{tabularx}
\end{table}

\section{Major Approaches to AV Safety}
\label{sec:result}

\par
In the following, we highlight four important classes of approaches to
AV safety---as well applicable to any other type of systems---together
with some notes offered to be taken away from the workshop
contributions and our
discussions~\cite{Gleirscher2018-SCAV-proceedings}.

When speaking of measures taken for safety assurance, some like to
use the term ``product-based'' to refer to the \emph{system under
  development} as the object to be delivered to a customer, and
``process-based'' to refer to the \emph{engineering process} or life
cycle of this system.  We will align the following discussion with
this nomenclature.

\subsection{Guaranteeing Reachability and Invariance}
\label{sec:safebyreach}

Here, we view AV safety as the problem of showing for a controlled
continuous dynamical physical system or a discrete state transition
system that, given an initial set,
\begin{enumerate}
\item a desired final set is always reached, and
\item a desired set (an invariant) will never be left,
\end{enumerate}
through appropriate influence of the controller or decision procedure and,
usually, under some additional constraints and the influence of
disturbances.  The properties 1 and 2 describe what is also called
\textbf{control stability}.  \emph{Safety measures} taken are, in a
strong sense, ``product-based'' and focus the control application.

From our discussions, we like to mention some state-of-the-art
practices and challenges to be solved with good generality:

\paragraph{The Scaling of Reachability Algorithms and Increasing
  Verification Coverage}
Research addresses techniques for set over- and under-approximation to
reduce ``flowpipe''\footnote{Approximation of a continuous dynamics
  given an initial set.} computation effort.
Discretization schemes based on hyper-reals or hyper-dense domains
help reducing verification to decidable problems.
However, the elaborate example of \emph{multi-lane traffic snapshots}
showed us, how hard it is to efficiently cover the space of real-world
scenarios coining initial sets and to keep the number of necessary
reachability checks low.

\paragraph{The Specification of Controller Performance} In distributed
control or with control technology relying on network communication,
research is investigating the derivation of lowest bounds of
acceptable computational performance (e.g.\ communication
requirements) from given control loop characteristics.

\paragraph{The Reduction of Model Uncertainty and Increasing
  Verification Confidence} We find it crucial to strengthen
discussions on
\begin{itemize}
\item %
  identifying uncertainties in models,
\item %
  identifying validity errors in models (i.e.~errors
  other than syntactic or semantic inconsistencies), and
\item %
  fixing such errors and assuring sufficient model accuracy.
\end{itemize}
Model parameters are often conservatively estimated by domain experts
(e.g.~parameters in probabilistic models).  Because modeling large
systems is challenging, relying solely on expert knowledge of such
parameters is not a first class approach to reducing model uncertainty
and, hence, can drastically decrease the confidence of any
verification results.  Clearly, on the other hand, the determination
of valid, ideally non-conservative, probabilities of hardware/software
failures and similar events is still far from being a trivial issue.

\paragraph{The Transfer of Verification Results to Realistic Settings}
\emph{Open and flexible simulation platforms} for autonomous systems
experimentation have shown to be a promising way for this.
Particularly, SCAV'18 authors have discussed controller code synthesis
from Matlab Simulink\furl{https://www.mathworks.com} to
the ROS\furl{http://www.ros.org} platform and simulation in ROS' RViz
environment and the simulation engine
Gazebo.\furl{http://gazebosim.org}
It is not unusual to distinguish between three increasingly more
realistic \textbf{stages of testing by simulation}:
\begin{itemize}
\item model-in-the-loop~(MIL), 
\item software-in-the-loop~(SIL), and
\item hardware-in-the-loop~(HIL).
\end{itemize}
Because the assumptions made by these ``loop simulations'' are
increasingly more realistic and, typically, increasingly more
expensive, it is important for test and verification engineers to
identify the earliest of these stages at which critical aspects of an
AV control application can be effectively tested or verified.  It is
well-known that late defect discovery is strongly positively
correlated with high costs of defect removal.

\subsection{Fault-Avoidance and Fault-Tolerance}
\label{sec:safebyfaulttol}

Here, we view AV safety as the problem of showing for a
\textbf{controller architecture, design, or implementation} that it
fulfills a number of, ideally quantitative, constraints referred to as
dependability (incl.\ reliability) and security requirements.
The measures taken are typically ``product-based'' and focus on
controller technology.
In practice, many of these constraints are either difficult to
identify, quantify, or specify; needless to say that their fulfillment
is very difficult in its own right.

Although, our discussions to that extent have been less extensive than
in
SCAV'17~\cite{Gleirscher2017-SCAV-report,Gleirscher2017-SCAV-proceedings},
we like to mention a classical issue investigated in a new technical
setting in one of the contributions:
For mixed-criticality applications using Ethernet technology and
software-defined networking~(SDN), it was found that
\textbf{acceptable fail-over performance in network reconfiguration}
is achievable.  Our discussion gives reason to believe that recent
developments in Ethernet technology may be a suitable option for
harmonizing currently heterogeneous and, hence, complex in-vehicle
networks.

An AV design determines the ``items'' that, across the disciplines
involved in AV engineering, define the \emph{contexts} of verification
tasks.  For example, \emph{functional safety} pertains to electrical,
electronic, and software parts of an AV.  It is well-known that the
decomposition of a system into items correlates with the structure of
the organization developing this system.  Undesired side-effects of
this phenomenon lead to gaps in the reasoning how local results from,
for example, functional safety improve the overall safety of a control
application.  The inverse problem arises when deciding about how
a constraint characterizing the control application has to be
distributed across parts of a controller implementation.

\subsection{Application of Normative Frameworks}
\label{sec:safebystd}

Here, we view AV safety as the problem of showing that certain
\textbf{guidelines, policies, standards, or regulations} (presumably
accepted by the corresponding stakeholders and regulatory authorities)
have been followed in an engineering process and that
\textbf{compliance} of this process with such a normative framework is
the governing part in the assurance argument.  Measures taken are, hence,
``process-based,'' focusing the controller engineering process.

Recent %
normative frameworks have shown deficiencies for some applications
that may have a dangerous impact on compliant products as sold in the
markets:
\begin{itemize}
\item Fail-operational concepts have been neglected in favor of
  fail-silent assumptions compatible with more traditional
  human-in-the-loop settings.
\item Agile engineering practices have been postponed in favor of
  rigid, waterfall-oriented
  processes, such as the V-model.
\item Open source components can not yet be safely integrated with
  proprietary embedded platforms.
\end{itemize}

However, in their applied forms, standards often lack clarity~(i.e.\
undesirable universality) in some places and over-specification~(i.e.\
undesirable restriction) in other places.  This circumstance deems
them less helpful as a quality control mechanism and in liability and
accountability cases.
Late availability of the frameworks postpones evaluations of how the
mentioned deficiencies are mitigated and, more severely, whether the
frameworks comply with the state of the science.

While some frameworks are applicable to very specific types
of systems (e.g.~four-\-wheel road vehicles), recent versions 
try to harmonize regulations and transfer them to systems similar
from the viewpoint of functional safety, such as motor cycles.
For example, with ISO 26262~(version 2), we were unable to identify
the relation of unintended behavior and failure.  Most likely, these
concepts are intertwined.  Our discussion indicates that safety
assurance against failures and unintended behavior (also summarized as
\emph{malfunctioning behavior}) and safety assurance of nominal
behavior should be accomplished in a single coherent framework to
identify subtle interference.  To this end, it was unclear whether
the \emph{safety of the intended function}~(SOTIF) standard will
address any of the class of approaches according to
\Cref{sec:safebyreach}.

\subsection{Case-based Argumentation}
\label{sec:safebyargument}

Here, we view AV safety as the problem of constructing \textbf{an
  individual argument} that the deployed controller is sufficiently
free of hazards or, similarly, that the controller is acceptably safe
with respect to the identified hazards.  This individual argument
results from \textbf{case-based explicit inductive or deductive
  reasoning} from evidence towards a desirable claim, or vice versa.
The term \emph{evidence} refers to measures of any kind taken to
substantiate \textbf{or refute} such a claim.  The adjective
``explicit'' is often understood as the hierarchical visualization of
this argument.  This visualization can, for example, be accomplished
by using the Goal Structuring
Notation~(GSN).\furl{http://www.goalstructuringnotation.info} Measures
taken can be ``product-'' or ``process-based'' and focus both the
control application and the controller implementation.

We like to mention an important issue pointed out in our discussion:
\emph{Safety and, more generally, assurance cases}, typically,
represent \emph{judgments} of expert committees about the most
probable effects of the taken safety measures.  Although such
judgments are structured by the argument, ``judgmental''
\emph{uncertainty} can well be injected through the evidence and
reasoning strategies used to construct the argument.  ``There is no
guarantee'' or, in other words, case-based arguments also bear the
tremendous challenge of delivering an acceptable level of confidence.

\section{Combining these Approaches}
\label{sec:synthesis}

For sake of brevity, we decided not to include references in this
report.  It is worth noting, the body of relevant literature on the
presented approaches is overwhelmingly large.

Anyway, the knowledgeable reader might find it quite easy to see that
these approaches to AV safety are related.  However, the character of
these relations and their implications are far less obvious.
In summary, the mentioned approaches more or less address two major
objectives characterizing the field:

\paragraph{Assurance of %
  Correctness} The approaches described in the
\Cref{sec:safebyreach,sec:safebyfaulttol,sec:safebyargument} provide
capabilities to accomplish the technical part of safe AV control, that
is, provide sufficient or compelling evidence that an AV controller
implementation actually fulfills its specification, not necessarily
questioning the acceptability or validity of this specification.  Such
a specification can be seen as a model and such a model, in most of
the practical cases, only captures part of the reality in its
assumptions.  This way, we again refer to the well-known fact that
``absolute safety'' cannot be guaranteed.

\paragraph{Assurance of Societal Agreeableness} %
Importantly, in addition to the first objective, the approaches
described in the \Cref{sec:safebystd,sec:safebyargument} provide
capabilities for the interaction of system vendors, regulatory
authorities, and scientific institutions with public society.  A
result of such an interaction can, for example, be an agreed
definition and \emph{specification} of what it means for an AV
controller to be ``acceptably safe'' or, in other words, of what it
means for safety risks in AV control to be ``as low as reasonably
practicable''%
\footnote{This is referred to as the ALARP principle~\cite{HSE2001}.}
in a given operational context.

The workshop talks and our discussions highlighted important steps
towards effective combinations of the approaches mentioned in the
\Cref{sec:safebyreach,sec:safebyfaulttol,sec:safebystd,sec:safebyargument}:
\begin{itemize}
\item %
  We have discussed the \textbf{integration of approximations of
    system dynamics with fault models} of AV controller
  implementations.
  While such approaches are very promising, we believe
  they will only be of practical use if they can clearly show how to deal with
  complex fault domains,
  moreover, how they can effectively relate these domains with the approximated
  physical loop dynamics.
\item %
  Regarding simulation platforms used in practice, we point to the
  challenge of \textbf{combining test and simulation with formal
    verification} and vice versa.
\item %
  We crossed the topic of \textbf{complexity and real-time
    performance} crucial for on-line applications of the discussed
  verification algorithms.
\end{itemize}

\vspace{-2em}

\paragraph{Conclusion}

Our empirical insights and feedback from interviews suggest that
practitioners in charge of AV control assurance need to be equipped
with and educated in stronger methods than they are currently applying
to challenge \textbf{assurance scalability}.  In conclusion, a
\textbf{sound combination of the aforementioned approaches} can be
seen as a grand challenge and a necessity if we like to push AV
control, in particular, and autonomous systems control, in general, to
a degree of safety desirable and acceptable in public and domestic
spaces.  \textbf{Automated and integrated formal methods} can play a
central role in this context.

\bibliographystyle{abbrv}
\bibliography{bibliography}
\end{document}